# The CMS Electromagnetic Calorimeter at the LHC

D.J.A. Cockerill, on behalf of the CMS ECAL Group
*Rutherford Appleton Laboratory, Chilton, Didcot, Oxfordshire, OX11 0QX, UK*

The CMS detector at the LHC is in the final stages of preparation. The high resolution Electromagnetic Calorimeter, which consists of nearly 76000 lead tungstate crystals, will play a crucial role in the coming physics searches undertaken by CMS. The design, status, and initial performance of the calorimeter, in test beams and with cosmic rays, will be reviewed.

## 1. INTRODUCTION

The Compact Muon Solenoid (CMS) detector [1] is a general purpose detector located at the LHC at CERN. The detector has been designed to be sensitive to a wide range of possible new physics, most notably for the detection of the decay products of the Higgs boson and for the detection of new particles predicted by Supersymmetry. For neutral Higgs bosons, with masses below ~140 GeV, the decay into two photons offers one of the cleanest channels for discovery [2]. This has led to the choice of a high resolution electromagnetic calorimeter (ECAL) [1, 3] for CMS, sited within the 3.8T solenoid and hadron calorimeter, with a target energy resolution of 0.5% for electrons and photons above 100 GeV.

The ECAL is a hermetic homogeneous calorimeter comprising 61200 lead tungstate ($PbWO_4$) crystals in the ECAL Barrel (EB), from $0 < |\eta| < 1.48$, and 14648 crystals in the ECAL Endcaps (EE) from $1.48 < |\eta| < 3$. The EB is subdivided into two halves, each of which contains 18 Supermodules (SMs). Each Endcap is divided into 2 'Dees'. The total crystal mass in the EB and EE is 67.4t and 22.9t respectively.

In front of each ECAL Endcap is a preshower detector, from $1.65 < |\eta| < 2.6$. The detector comprises two orthogonal planes of silicon strip detectors behind $2X_0$ and $1X_0$ of lead absorber respectively, in order to discriminate against $\pi^o$s which form part of the background to photons from Higgs decays. The silicon detectors have 32 strips, each 63mm long, on a pitch of 1.9mm. The total silicon area for both endcaps is $16.5m^2$, and comprises 4300 detectors with $1.4.10^5$ channels. It will be installed in CMS towards the end of 2008.

## 2. CRYSTALS AND PHOTODETECTORS

Lead tungstate crystals have been chosen for the ECAL to provide a homogeneous scintillating medium for optimal energy resolution. They have important properties which are crucial for use at the LHC. The crystals are fast emitters of scintillation light with 80% of the light released within 25nsec. The crystals have a short radiation length of 0.89cm and a small Moliere radius of 2.2cm which leads to a compact calorimeter with good granularity. The scintillation light has an emission peak at 425nm which is well suited for coupling to available photo-detectors. The scintillation emission has a temperature dependence of $-2\%/^oC$. The relatively low light yield (1.3% of NaI) requires the use of photo-detectors with gain in the presence of the 3.8T field of CMS.

The crystals are radiation resistant up to very high integrated doses. However, under irradiation at room temperature, colour centres form and self anneal, causing the light output, and consequently the signal seen in the photo-detector, to vary with dose rate. These changes are followed and corrected using a precise crystal monitoring system using laser light which is sent to each crystal. Ionising radiation has not been observed to affect the scintillation mechanism in PbWO4 crystals.

All crystals are slightly tapered. This provides an off-pointing angle of ~3$^o$ with respect to the interaction vertex in order to optimize detector hermeticity. The 34 types of EB crystals are 23cm long (25.8 Xo) by ~2.6x2.6cm$^2$ at the rear. All EE crystals are identical and are 22cm (24.7 Xo) long by 3x3cm$^2$ at the rear. The area, in $\Delta\eta\times\Delta\phi$, subtended by each crystal in the EB is 0.0175x0.0175, while in the EE it varies from 0.0175x0.0175 to 0.05x0.05. The EB employs avalanche photodiodes (APDs) to detect the scintillation light. Two 5x5mm$^2$ APDs are glued to each crystal. They have a quantum efficiency of ~75% and are operated at a gain of 50. The APDs have a gain dependence with temperature of -2.4%/$^o$C. Since the crystal scintillation has a temperature dependence of similar magnitude, and with the same sign, the ECAL is operated to within ±0.05$^o$C to minimize temperature effects. The EE employs Vacuum Photo-triodes (VPTs) which are more radiation resistant than silicon diodes. A single VPT is glued to each crystal. The devices are 26.5mm in diameter, with a UV glass window with an active area of 280mm$^2$. The VPTs have gains of ~8-10 in the 3.8T field and a quantum efficiency of ~20% at 420nm.

## 3. READOUT AND COMMISIONING

The on detector readout is organized around units of 25 (5x5) crystals. The APDs and VPTs are connected to multi gain preamplifiers (MGPAs) each of which has 3 gain ranges. Each of the 3 ranges is fed to a 12 bit ADC (one of 4 ADCs in the AD41240 chip) which digitizes the data at 40MHz. Embedded logic selects the highest non-saturated gain. The full dynamic ranges in the EB and EE are 60pC (~1.5TeV) and 12.8pC (~1.6 – 3.0 TeV) respectively. The ADCs are connected to a frontend (FE) readout card which stores the data in a pipeline (256x25nsec), until reception of a level one (L1) trigger accept from CMS, and performs transverse energy sums (also called trigger primitive sums) every 25nsec. The sums are transmitted to a Trigger Concentrator Card (TCC) in the counting room. The TCCs receive the trigger primitive sums and send complete trigger tower energy sums to the Regional Calorimeter Trigger (RCT) at 40MHz.

The FE card transmits the data to a Data Concentrator Card (DCC) in the counting room. The DCC also reads the TCC information, upon the L1 accept, and performs data reduction and data transfer to the CMS data acquisition system. Data communication between the FE card and the counting room is carried out through custom fibre optic links operating at 800MHz. The on-detector electronics are fabricated in 0.25µm CMOS for good radiation tolerance.

The complete EB has been installed in CMS since July 2007 and the complete EE since August 2008. Both detectors have been fully integrated into the readout chain of CMS. The EB has been fully integrated into the trigger readout chain and has participated in several months of CMS cosmic runs, recording millions of cosmic ray events (see Fig. 1, left). The commissioning has been extremely important for debugging the trigger and data paths and for timing the trigger primitives. CMS is now able to trigger with the full EB.

Detector timing has been carried out between EB, EE, and CMS using muon bremstrahlung events which cross the Muon and ECAL detectors. Further significant improvements in timing have been made using the LHC beam dump events in September 2008. The events, from the dumps of ~2.10$^9$ protons onto a tungsten collimator 150m upstream of CMS, produce ~10$^9$ simultaneous particles which cross the CMS detector. Such an event is shown in Fig. 1 (right).

34<sup>th</sup> International Conference on High Energy Physics, Philadelphia, 2008

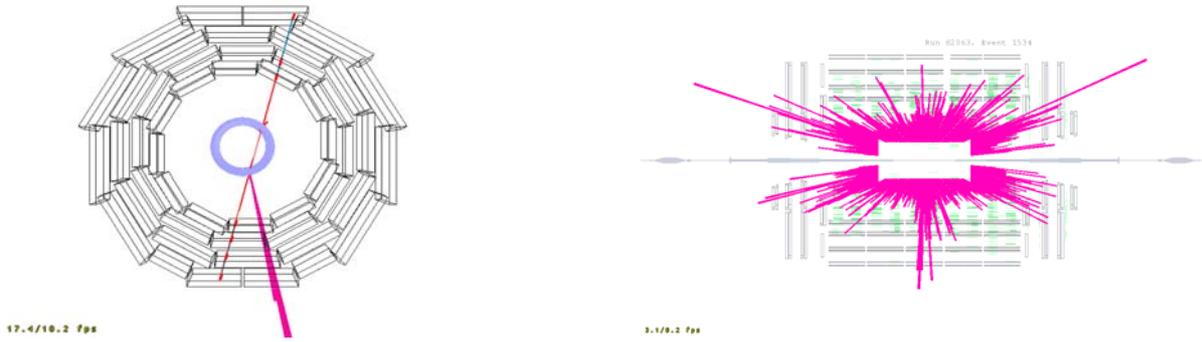

Figure 1: (left) a cosmic ray event with bremstrahlung in the EB, (right) ECAL display from an LHC beam dump event

## 4. ENERGY RESOLUTION AND CALIBRATION

Figure 2 (left) shows the energy resolution as a function of energy for electrons which were centrally incident (4x4mm$^2$) on 25 individual crystals in a 5x5 SM readout unit. The reconstructed energy for each incident electron was calculated from the sum of energy deposits over the 3x3 array centred on the struck crystal using common inter-calibration constants. The resolution is better than 0.5%, above 100GeV, meeting the ECAL performance targets.

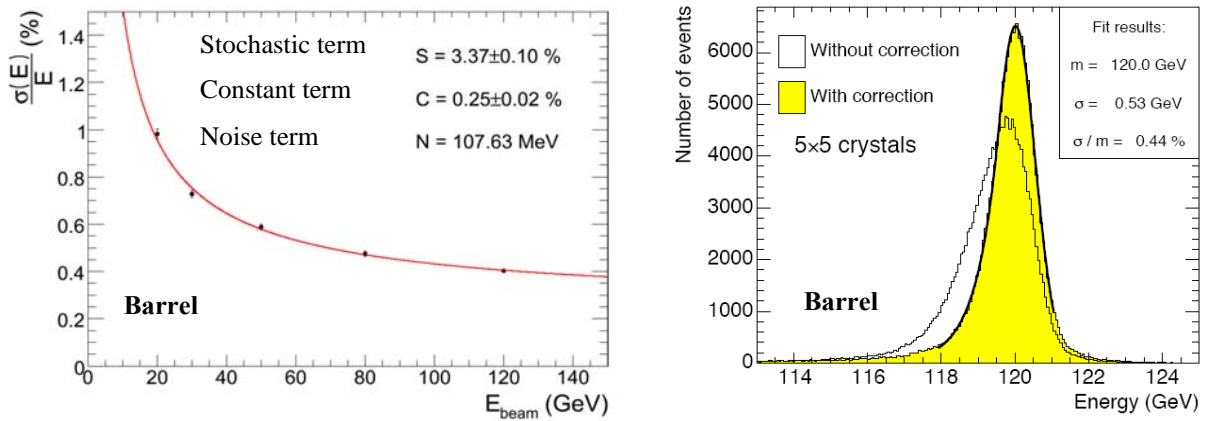

Figure 2: (left) the energy resolution as a function of energy for electrons centrally (4x4mm$^2$) incident on crystals, (right) the energy resolution at 120 GeV for incident electrons over an area of 20x20mm$^2$.

Figure 2 (right) shows the energy resolution at 120 GeV for incident electrons over an area of 20x20mm$^2$ (~80% of the crystal front face) before (unshaded) and after (shaded) the application of a universal position correction function to the reconstructed energy. The reconstructed energies were calculated from the sum of energy deposits over the 5x5 array centred on the struck crystal. The energy resolution after position correction is 0.44% at 120GeV.

All 36 SMs were exposed to cosmic rays in order to obtain pre-calibration coefficients for CMS. 9 SMs (1/4 of EB) were also exposed to test beam electrons. The pre-calibration coefficients obtained with the test beam were measured to a precision of 0.3%. The coefficients were compared to those obtained with cosmic ray data as shown in Fig. 3 (left). The coefficients agree to within ~1.5%, averaged over all the crystals. This sets the level of pre-calibration precision to be expected for the remaining 27 SMs which were only exposed to cosmic rays. A comparison of beam data from 460 EE crystals, to laboratory light yield measurements on crystals and yield measurements on VPTs, indicates that the EE pre-calibration precision will be ~9% RMS.

Fast in-situ calibration procedures are being prepared for startup at the LHC. The procedures will use the $\phi$ symmetry of energy deposits in the detector and calibrations with $\pi^o$s to improve the inter-calibration precision, particularly for the EE. The EB pre-calibration will be exploited for validation and tuning. A SM was exposed to $\pi^o$s at a test beam as a preliminary exercise to evaluate the possibilities of $\pi^o$ calibration. Figure 3 (right) shows the invariant mass measurements for photon pairs coming from $\pi^o$s generated through charge exchange with a beam of $\pi^-$ particles incident on a metal target. The width of the mass peak is 5.5%.

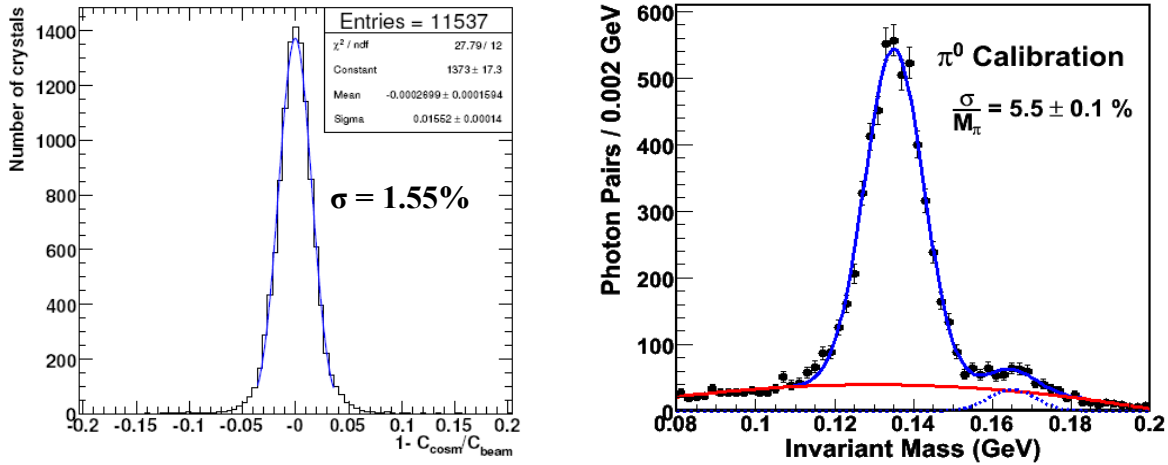

Figure 3: (left) a comparison of the inter-calibration coefficients obtained from cosmic rays with respect to those obtained with test beam electrons for 7 Supermodules, (right) a reconstructed $\pi^o$ resonance in a Supermodule exposed to $\pi^o$s from a test beam.

## 5. CONCLUSIONS

The high resolution CMS crystal calorimeter has been successfully installed and commissioned. The full EB trigger is ready and operational. The energy resolution has been measured in test beams and meets the 0.5% design target above 100 GeV. Extensive pre-calibration has been carried out, to a precision of 0.3-1.55% in the EB and 9% in the EE. Fast in-situ calibration procedures are being prepared for LHC running. The full calorimeter was used during the first beam dump events at the LHC in September 2008, providing important timing and performance data.

### Acknowledgments

I would like to thank the CMS ECAL collaboration and the conference organizers for the opportunity to give this talk and also to acknowledge and thank the Science and Technology Facilities Council, UK, and the Particle Physics Department, Rutherford Appleton Laboratory, UK for the funding to attend this excellent conference.

### References

[1] The CMS experiment at the CERN LHC, The CMS Collaboration, S Chatrchyan *et al*, 2008 JINST **3** S08004

[2] Physics TDR Volume II, 'Physics Performance', The CMS Collaboration, 2007 *J. Phys. G: Nucl. Part. Phys.* **34** 995-1579

[3] Physics TDR Volume 1, 'Detector Performance and Software', The CMS Collaboration, CERN/LHCC 2006-001